# Transformation-based generalized spatial regression using the spmoran package: Case study examples

Daisuke Murakami

Department of Statistical Data Science, Institute of Statistical Mathematics, 10-3 Midori-cho, Tachikawa, Tokyo, 190-8562, Japan

Table of Contents

# 1. Introduction

## 1.1. Outline

This study presents application examples of generalized spatial regression modeling for count data and continuous non-Gaussian data using the spmoran package (version 0.2.2 onward). Section 2 introduces the model. The subsequent sections demonstrate applications of the model for disease mapping, spatial prediction and uncertainty modeling, and hedonic analysis.

The R codes used in this vignette are available from https://github.com/dmuraka/spmoran. Another vignette focusing on Gaussian spatial regression modeling is also available from the same GitHub page (and Murakami 2017).

## 1.2. Model

We consider the following generalized spatial regression model (see Murakami et al., 2021):

$$\varphi_{\boldsymbol{\theta}}(y_i) = z_i, \quad z_i = \sum_{k=1}^{K} x_{i,k}\beta_{i,k} + w_i + \varepsilon_i, \quad w_i \sim N(0, c(d_{ij})), \quad \varepsilon_i \sim N(0, \sigma^2), \tag{1}$$

where $\varphi_{\boldsymbol{\theta}}(\cdot)$ is a transformation function normalizing the $i$-th explained variable $y_i$. $x_{i,k}$ is the $k$-th explanatory variable, $\beta_{i,k}$ is a fixed or random coefficient, which may vary spatially and/or non-spatially (the distribution for $\beta_{i,k}$ is omitted from Eq. (1) for simplicity). $w_i$ is a term capturing residual spatial dependence. Moran eigenvectors, which are spatial basis functions, are used to model the spatially dependent processes in $\beta_{i,k}$ and $w_i$. The model may be rewritten as follows:

$$y_i = \varphi_{\boldsymbol{\theta}}^{-1}(z_i), \quad z_i = \sum_{k=1}^{K} x_{i,k}\beta_{i,k} + w_i + \varepsilon_i, \quad w_i \sim N(0, c(d_{ij})), \quad \varepsilon_i \sim N(0, \sigma^2). \tag{2}$$

Eq. (2) suggests that $y_i$ is assumed to have a distribution that is obtained by transforming a Gaussian distributed $z_i$ using the $\varphi_{\boldsymbol{\theta}}^{-1}(\cdot)$ function. This model describes a wide variety of non-Gaussian data including count data by flexibly specifying the transformation function.

The transformation function is defined by concatenating $D$ sub-transformation functions:

$$\varphi_{\boldsymbol{\theta}}(y_i) = \varphi_{\boldsymbol{\theta}_D}\left(\varphi_{\boldsymbol{\theta}_{D-1}}\left(\cdots \varphi_{\boldsymbol{\theta}_2}\left(\varphi_{\boldsymbol{\theta}_1}(y_i)\right)\cdots\right)\right), \tag{3}$$

where $\varphi_{\boldsymbol{\theta}_d}(\cdot)$ is the $d$-th sub-transformation function depending on a set of parameters $\boldsymbol{\theta}_d$. For continuous explained variables, the spmoran package provides the following specifications for $\varphi_{\boldsymbol{\theta}}(\cdot)$ (see Figure 1):

(a) For non-negative $y_i$, the Box-Cox transformation is available (left of Figure 1).
(b) For non-Gaussian $y_i$ (e.g., skew and fat-tail distribution), the SAL transformation Eq. (4) (Rios and Tobar, 2019), which is a non-linear transformation, is iterated $D$ times to accurately normalize $y_i$ (middle of Figure 1):

$$\varphi_{\boldsymbol{\theta}_d}(y_i) = \theta_{d,1} + \theta_{d,2}\sinh(\theta_{d,3}\arcsinh(y_i) - \theta_{d,4}), \qquad (4)$$

where $\boldsymbol{\theta}_d \in \{\theta_{d,1}, \theta_{d,2}, \theta_{d,3}, \theta_{d,4}\}$.

(c) For non-negative and non-Gaussian $y_i$, the Box-Cox transformation is applied first, and the SAL transformation is iterated $D$ times after that to accurately normalized $y_i$ (right of Figure 1).

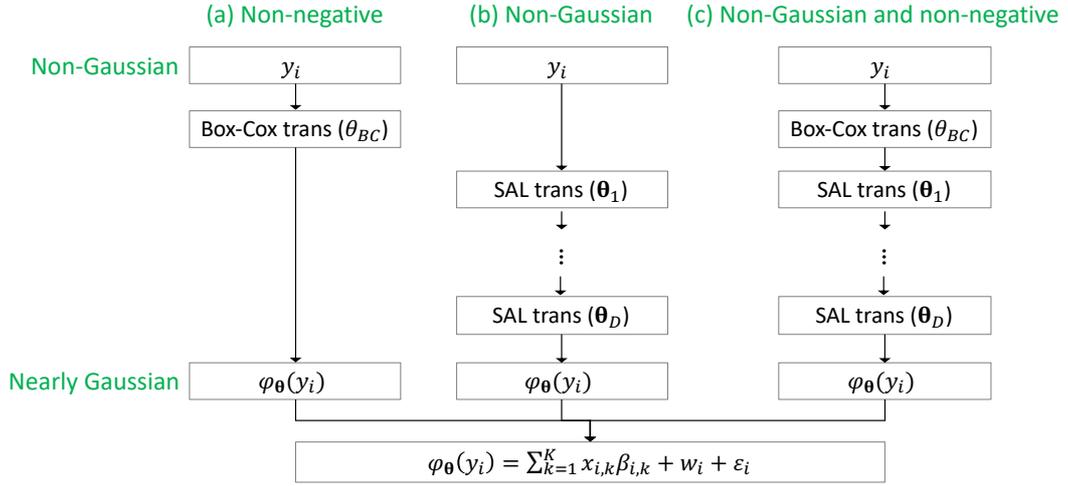

Figure 1: Transformation functions for continuous variables

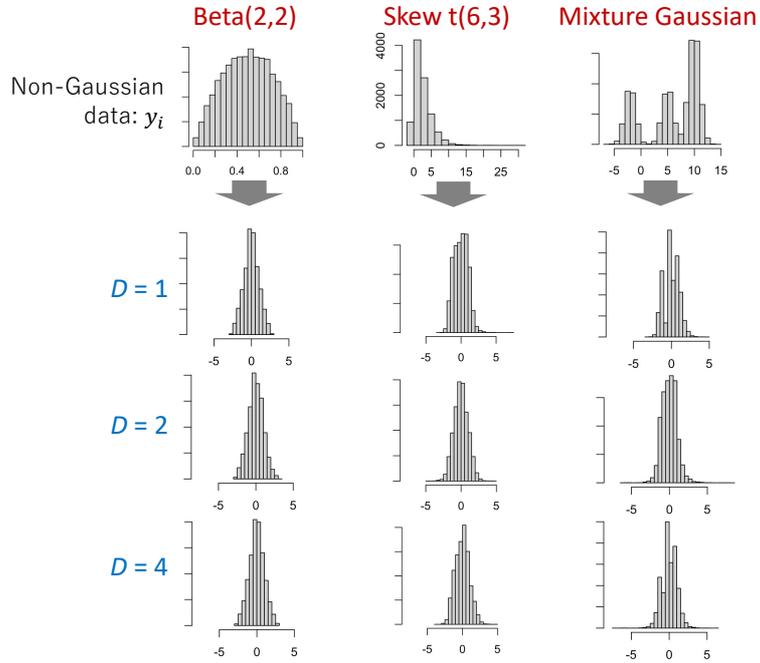

Figure 2: Results of applying the iterative SAL transformations to simulated data generated from beta, skew t, and Gaussian mixture distributions. The top three panels represent histograms of the simulated non-Gaussian data, and the bottom nine panels show the histograms after the transformations. $D$ is the number of transformations.

As illustrated in Figure 2, the iteration of the SAL transformations converts a wide variety of non-Gaussian data $y_i$ to Gaussian data $\varphi_\theta(y_i)$ quite flexibly. Thus, the generalized regression model Eq. (1) is available for a wide variety of non-Gaussian data.

This model Eq. (1) is also available for count data by applying a (log-)Gaussian transformation approximating a count data distribution. In the spmoran package, the following transformations are implemented:

(d) For (over-dispersed) Poisson counts, a log-Gaussian approximation proposed by Murakami and Matsui (2021) is available (left of Figure 3). Based on them, accuracy of the approximate model is almost the same as the conventional over-dispersed Poisson regression.

(e) For counts which do not obey the Poisson distribution, the log-Gaussian approximation is applied first to roughly normalize the data, and the SAL transformation is iterated after that to identify the most likely distribution (i.e., probability mass function) (right of Figure 3).

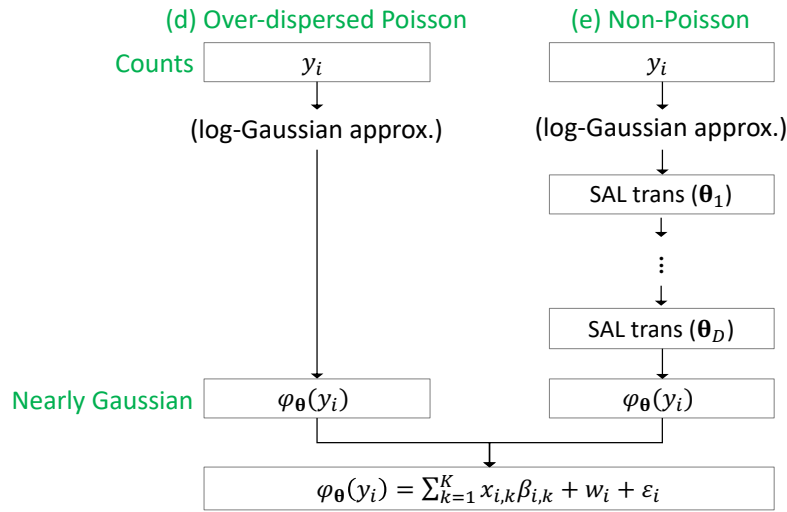

Figure 3: Transformation functions for count variables

## 1.3. Coding for specifying the transformation

In the spmoran package, the transformation function $\varphi_{\boldsymbol{\theta}}(\cdot)$ in Eq. (1) is specified by using the nongauss_y function. Here is a code (blue part) to specify (a) for non-negative $y_i$:

```
> ng_a     <- nongauss_y(y_nonneg=TRUE)
Box-cox transformation f() is applied to y to estimate
y ~ P( xb, par )   (or f(y,par)~N(xb, sig) )

 - P(): Distribution estimated through the transformation
 - xb : Regression term with fixed and random coefficients in b
        which is specified by resf or resf_vc function
 - par: Parameter estimating data distribution
```

y_nonneg = TRUE constraints the explained variables to not to have negative values. The output from the nongauss_y function is used as an input of the resf or resf_vc function to estimate Eq. (1). The transformation (b) for non-Gaussian $y_i$ and (c) for non-negative and non-Gaussian $y_i$ are specified as follows (D = 2 is assumed):

```
> ng_b     <- nongauss_y(tr_num=2)
2 SAL transformations are applied to y to estimate
y ~ P( xb, par )   (or f(y,par)~N(xb, sig) )

 - P(): Distribution estimated through the transformation(s)
 - xb : Regression term with fixed and random coefficients in b
        which is specified by resf or resf_vc function
 - par: Parameters estimating data distribution

> ng_c     <- nongauss_y(y_nonneg=TRUE,tr_num=2)
Box-Cox and 2 SAL transformations f() are applied to y to estimate
y ~ P( xb, par )   (or f(y,par)~N(xb, sig) )

 - P(): Distribution estimated through the transformation(s)
 - xb : Regression term with fixed and random coefficients in b
        which is specified by resf or resf_vc function
 - par: Parameters estimating data distribution
```

where tr_num (=D) specifies the number of SAL transformations. Finally, the transformation (d) for over-dispersed Poisson counts and (e) for other counts are specified as follows:

```
> ng_d     <- nongauss_y(y_type="count")
Log-Gaussian approximation estimating
y ~ oPois( mu, sig ), mu = exp( xb )

 - oPois(): Overdispersed Poisson distribution
 - xb     : Regression term with fixed and random coefficients in b
            which is specified by resf or resf_vc function
 - sig    : Dispersion parameter (overdispersion if sig > 1)
```

```
> ng_e     <- nongauss_y(y_type="count",tr_num=2)
Log-Gaussian and 2 SAL transformations are applied to y to estimate
 y ~ P( mu, par ), mu = exp( xb )

 - P(): Distribution estimated through the transformations
 - xb : Regression term with fixed and random coefficients in b
        which is specified by resf or resf_vc function
 - par: Parameters estimating data distribution
```

where y_type specifies data type ("count" for count variables and "continuous" for continuous variables (default)).

The subsequent sections present application examples of the model for count data (Section 2) and continuous data (Sections 3-4).

## 2. Example 1: Disease mapping and regression with count data

This section demonstrates a count regression modeling for epidemic data considering spatially varying coefficients, residual spatial dependence, and heterogeneity across years. The estimated model is used mainly for disease mapping and uncertainty modeling.

### 2.1. Data

This section uses sf, rgeos, CARBayesdata, spdep, spmoran packages:

```
> library(sf);library(rgeos);library(CARBayesdata);library(spdep);library(spmoran)
```

We employ the pollution-health data (pollutionhealthdata), which is available from the CARBayesdata package. The data consists of respiratory hospitalization data, air pollution, and covariate data for the Greater Glasgow (2007 - 2011) by 271 Intermediate Geographies (IG).

```
> data("pollutionhealthdata")
> head(pollutionhealthdata)
        IG year observed  expected     pm10 jsa price
1 S02000260 2007       97  98.24602 14.02699 2.25 1.150
2 S02000261 2007       15  45.26085 13.30402 0.60 1.640
3 S02000262 2007       49  92.36517 13.30402 0.95 1.750
4 S02000263 2007       44  72.55324 14.00985 0.35 2.385
5 S02000264 2007       68 125.41904 14.08074 0.80 1.645
6 S02000265 2007       24  55.04868 14.08884 1.25 1.760
```

Explained variable (y) is the number of hospitalization due to respiratory disease (observed). Explanatory variables (x) are the average particulate matter concentration (pm10), the percentage of working age people who are in receipt of Job Seekers Allowance, a benefit paid to unemployed people looking for work (jsa), and average property price (divided by 100,000) (price). Random effects by years are considered to estimate heterogeneity across years (xgroup). Besides, the expected numbers of hospitalizations based on Scotland-wide respiratory hospitalization rates (expected) is used as an offset variable. These variables are specified as follows:

```
> y      <- pollutionhealthdata[,"observed"]
> x      <- pollutionhealthdata[,c("jsa","price","pm10")]
> xgroup <- pollutionhealthdata[,"year"]
> offset <- pollutionhealthdata[,"expected"]
```

A binary contiguity matrix, which is generated from the spatial polygons by IGs (GGHB.IG), is used for modeling spatial dependence:

```
> data("GGHB.IG")
> W.nb   <- poly2nb(GGHB.IG)
> W.list <- nb2listw(W.nb, style = "B")
> W      <- nb2mat(W.nb, style = "B")
```

As explained, Moran eigenvectors are used to model spatially dependent process. Here is a code generating the eigenvectors from the W matrix:

```
> s_id   <- pollutionhealthdata[,"IG"]
> meig   <- meigen(cmat=W, s_id = s_id )
 109/271 eigen-pairs are extracted
```

where cmat specifies a spatial proximity matrix, and s_id specifies zone ID (the i-th row of cmat and the element of s_id that appears in the i-th are associated).

2.2. Model

This section considers two specifications for y. The former (ng1) assumes y to obey an over-dispersed Poisson distribution. The latter assumes a more general distribution, and estimates it through the SAL transformation (ng2):

```
> ng1     <- nongauss_y( y_type = "count")
Log-Gaussian approximation estimating
y ~ oPois( mu, sig ), mu = exp( xb )

 - oPois(): Overdispersed Poisson distribution
 - xb     : Regression term with fixed and random coefficients in b
            which is specified by resf or resf_vc function
 - sig    : Dispersion parameter (overdispersion if sig > 1)

> ng2     <- nongauss_y( y_type = "count", tr_num=1 )
Log-Gaussian and 1 SAL transformations are applied to y to estimate
 y ~ P( mu, par ), mu = exp( xb )

 - P(): Distribution estimated through the transformations
 - xb : Regression term with fixed and random coefficients in b
        which is specified by resf or resf_vc function
 - par: Parameters estimating data distribution
```

The outputs ng1 and ng2 are used as inputs for the resf or resf_cv function. The resf function estimates spatial regression models without spatially varying coefficients (SVCs) while the resf_vc function estimates models with SVCs (see Murakami, 2017). Here, we estimate the following models:

```
> mod1    <- resf(y=y, x=x, meig=meig, xgroup=xgroup,nongauss=ng1)
> mod2    <- resf(y=y, x=x, meig=meig, xgroup=xgroup,nongauss=ng2)
> mod3    <- resf_vc(y=y, x=x, xgroup=xgroup, offset=offset,meig=meig,nongauss=ng1)
> mod4    <- resf_vc(y=y, x=x, xgroup=xgroup, offset=offset,meig=meig,nongauss=ng2)
```

mod1 and mod2 assume constant coefficients while mod3 and mod4 assume SVCs on x. For the distribution of y, mod1 and mod3 assume an over-dispersed Poisson distribution while mod2 and mod3 adjust the distribution using the SAL transformation to identify the most likely distribution. The BIC values are -260.1 (mod1), -256.2 (mod2), -274.2 (mod3), and -271.7 (mod4). mod3, which is an over-dispersed Poisson SVC model, is selected as the best model. Note that the BIC is based on a Gaussian likelihood approximating the Poisson model, which is different from the conventional Poisson likelihood.

The estimation result of mod3 is as below. The intercept and coefficient on price are estimated spatially varying while the coefficients on jsa and pm10 are estimated constant. As shown in the bottom, the BIC of mod3 is considerably better than the BIC of the NULL model (74.9), which is a log-Gaussian model approximating the conventional Poisson regression:

```
> mod3
Call:
resf_vc(y = y, x = x, xgroup = xgroup, offset = offset, meig = meig,
    nongauss = ng1)

----Spatially varying coefficients on x (summary)----

Coefficient estimates:
  (Intercept)            jsa              price             pm10
 Min.   :-0.6504   Min.   :0.06149   Min.   :-0.33538   Min.   :0.02834
 1st Qu.:-0.5831   1st Qu.:0.06149   1st Qu.:-0.23431   1st Qu.:0.02834
 Median :-0.5526   Median :0.06149   Median :-0.19311   Median :0.02834
 Mean   :-0.5478   Mean   :0.06149   Mean   :-0.18184   Mean   :0.02834
 3rd Qu.:-0.5163   3rd Qu.:0.06149   3rd Qu.:-0.13469   3rd Qu.:0.02834
 Max.   :-0.3929   Max.   :0.06149   Max.   : 0.04439   Max.   :0.02834

Statistical significance:
                      Intercept  jsa price pm10
Not significant               0    0   205    0
Significant (10% level)       0    0    70    0
Significant ( 5% level)       0    0   180    0
Significant ( 1% level)    1355 1355   900 1355

----Variance parameters---------------------------------

Spatial effects (coefficients on x):
                   (Intercept) jsa      price pm10
random_SE           0.07496275   0 0.09383671    0
Moran.I/max(Moran.I) 0.72069442  NA 0.37600487   NA

Group effects:
             xgroup
ramdom_SE 0.1219861

----Estimated probability distribution of y--------------
                Estimates
skewness         1.026517
excess kurtosis  1.752394

----Error statistics------------------------------------
                                                stat
dispersion parameter                        3.132744
deviance explained (%)                     82.977533
Gaussian rlogLik approximating the model  173.152374
AIC                                       -326.304748
BIC                                       -274.189181

NULL model: glm( y ~ x, offset = log( offset ), family = poisson )
   Gaussian (r)loglik approximating the model: -19.4258
   ( AIC: 48.85159,  BIC: 74.90938 )
```

The estimated group effects are as follows:

```
> mod3$b_g
[[1]]
                 Estimate         SE    t_value
xgroup_2007   0.052882464 0.02678485   1.974343
xgroup_2008   0.107183516 0.02409724   4.447959
xgroup_2009   0.007175285 0.02767944   0.259228
xgroup_2010  -0.083975107 0.02474086  -3.394187
xgroup_2011  -0.083266159         NA         NA
```

While regression coefficients for the transformed y is often difficult to interpret, marginal effect $dy_i/dx_{i,k}$ which quantifies the magnitude of change in *i*-th explained variable ($y_i$) for one unit change in the *k*-th explanatory variable ($x_{i,k}$), can be evaluated using the coef_marginal function if the resf function is used while the coef_marginal_vc function if the resf_vc function is used:

```
> coef_marginal_vc(mod3)
Call:
coef_marginal_vc(mod = mod3)

----Marginal effects from x (dy_i/dx_i) (summary)----
 (Intercept)          jsa             price            pm10
 Mode:logical   Min.   : 1.333   Min.   :-34.568   Min.   :0.6144
 NA's:1355      1st Qu.: 3.584   1st Qu.:-17.135   1st Qu.:1.6520
                Median : 4.652   Median :-12.722   Median :2.1441
                Mean   : 4.915   Mean   :-13.342   Mean   :2.2654
                3rd Qu.: 5.979   3rd Qu.: -9.379   3rd Qu.:2.7556
                Max.   :11.795   Max.   :  7.291   Max.   :5.4363

Note: Medians are recommended summary statistics
```

For example, the median of pm10 suggests that the number of hospitalizations increases 2.1441 for every 1.0 increase of pm10.

The explained variables and the predicted values are plotted below. This result confirms accuracy of the model:

```
> plot(y,mod3$pred[,1])
```

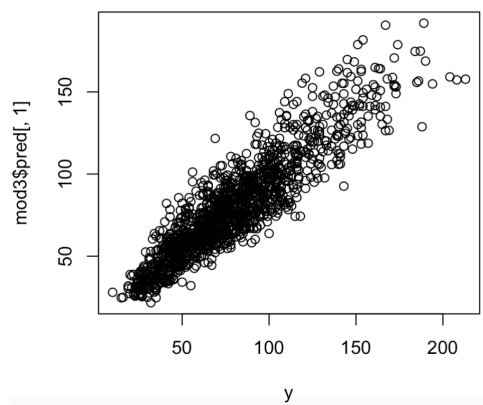

In addition to the predicted values plotted above, the resf and resf_vc functions return quantiles of the predicted values, which are estimated based on the modeled probability density/mass function. They are displayed as follows:

```
> mod3$pred_quantile[1:2,]
     q0.01   q0.025    q0.05    q0.1     q0.2     q0.3     q0.4     q0.5     q0.6
1 52.05107 56.02032 59.67535 64.18632 70.10760 74.71338 78.88783 83.00021 87.32698
2 16.12654 17.23963 18.25821 19.50748 21.13521 22.39256 23.52602 24.63725 25.80097
      q0.7     q0.8      q0.9    q0.95    q0.975    q0.99
1 92.20619 98.26375 107.32872 115.4419 122.97388 132.35148
2 27.10695 28.71957  31.11597  33.2450  35.20924  37.63946
```

The quantiles are useful for evaluating uncertainty in disease mapping (see below).

2.3. Regression and disease mapping

The predicted values are available for disease mapping. Here, we consider mapping the patterns in 2007. Here is a code to create a dataset including observed counts in 2007 (obs), predicted counts and their standard errors (pred), estimated varying coefficients (b_est), and quantiles of the predicted values (pred_qt), and convert the dataset to sf format, which is a spatial data format, for mapping:

```
> obs    <- y[pollutionhealthdata[,"year"] == 2007]
> pred   <- mod3$pred[pollutionhealthdata[,"year"] == 2007, ]
> b_est  <- mod3$b_vc[pollutionhealthdata[,"year"] == 2007,]
> pred_qt<- mod3$pred_quantile[pollutionhealthdata[,"year"] == 2007,]
>
> poly   <- st_as_sf(GGHB.IG)
> poly   <- cbind(poly, obs, pred, b_est, pred_qt)
```

The predicted counts are as mapped together with the observed counts below. The result suggests that the estimated model accurately identifies the spatial pattern underlying the respiratory disease.

```
> plot(poly[,c("obs","pred")],axes=TRUE, lwd=0.1, key.pos = 1)
```

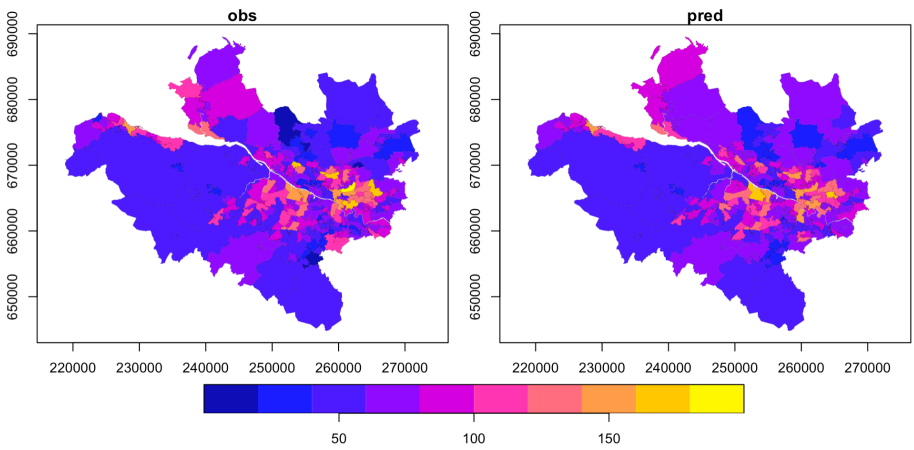

Here is a code to map percentile (0.025, 0.50, 0.975%) of the predicted values. This map suggests higher uncertainty in the central urban area while lower uncertainty in the suburban areas.

```
> plot(poly[,c("q0.025","q0.5","q0.975")],axes=TRUE, lwd=0.1, key.pos = 1)
```

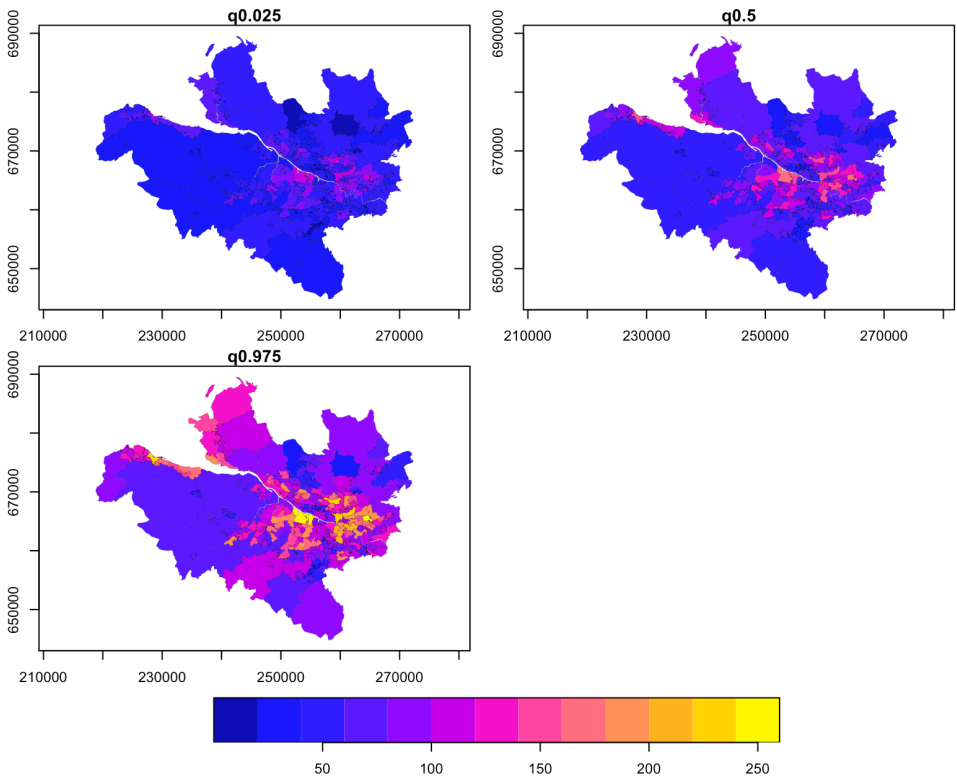

Finally, the estimated spatially varying intercept and coefficients on price are plotted below:

```
> plot(poly[,"X.Intercept."],axes=TRUE,lwd=0.1, key.pos = 1)
```

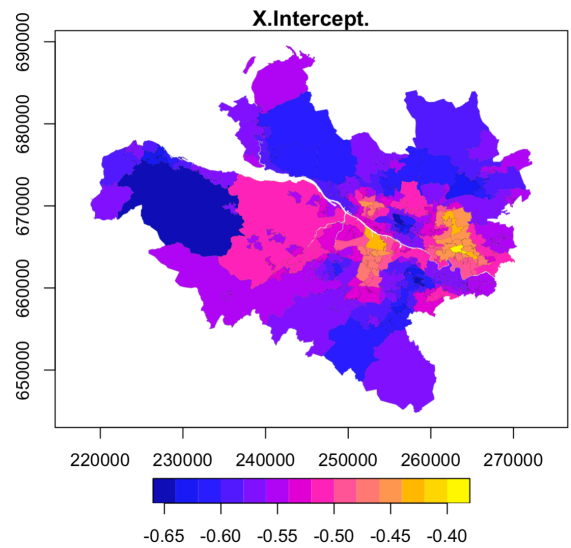

```
> plot(poly[,"price"],axes=TRUE,lwd=0.1, key.pos = 1)
```

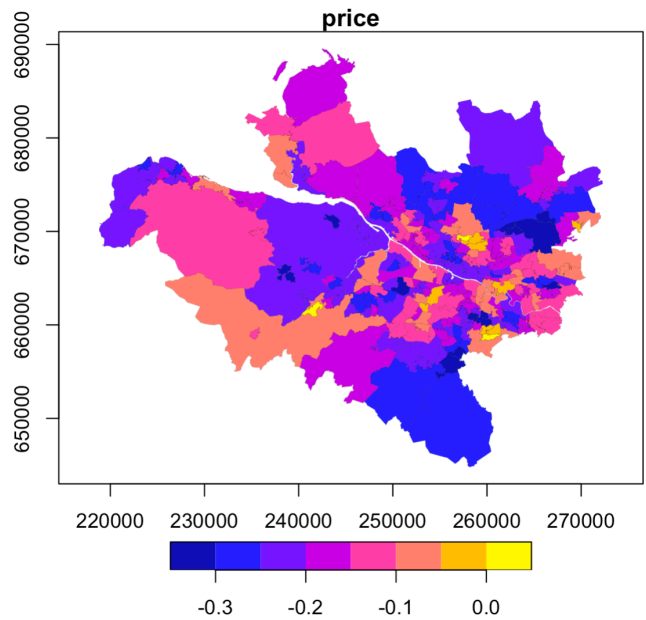

# 3. Example 2: Spatial prediction and uncertainty analysis for non-Gaussian data

This section demonstrates a non-Gaussian spatial regression modeling for spatial interpolation and uncertainty modeling.

## 3.1. Data

This section uses the sf, automap, and spmoran packages:

```
> library(sf);library(automap);library(spmoran)
```

The meuse data, which we will use in this section, consists of heavy metal concentrations (cadmium, copper, lead, zinc) measured in a flood plain along the river Meuse and explanatory variates:

```
> data(meuse)
> meuse[1:5,]
       x      y cadmium copper lead zinc  elev       dist   om ffreq soil lime landuse dist.m
1 181072 333611    11.7     85  299 1022 7.909 0.00135803 13.6     1    1    1      Ah     50
2 181025 333558     8.6     81  277 1141 6.983 0.01222430 14.0     1    1    1      Ah     30
3 181165 333537     6.5     68  199  640 7.800 0.10302900 13.0     1    1    1      Ah    150
4 181298 333484     2.6     81  116  257 7.655 0.19009400  8.0     1    2    0      Ga    270
5 181307 333330     2.8     48  117  269 7.480 0.27709000  8.7     1    2    0      Ah    380
```

We analyze the zinc concentration in ppm (zinc). As shown in the histogram below, the zinc data does not have a Gaussian distribution:

```
> y    <-meuse$zinc
> hist(y)
```

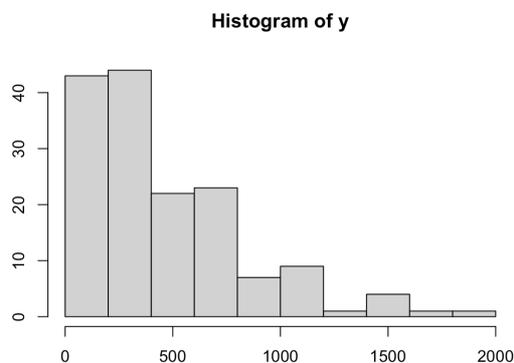

Histogram of y

Here is the spatial plot of the zinc concentration:

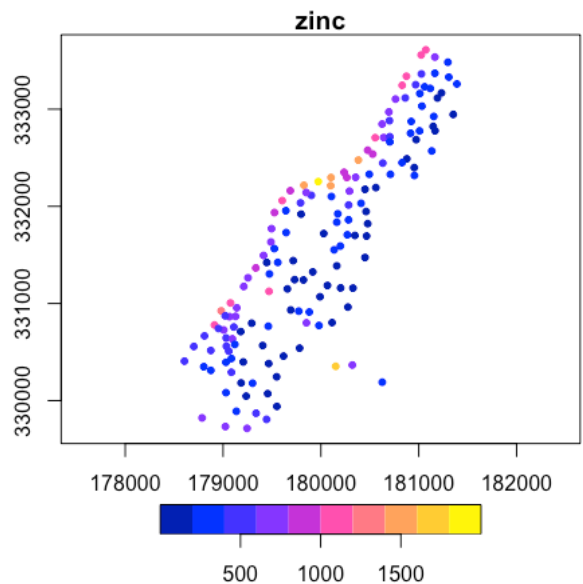

We use dist (distance to river Meuse), ffreq2 (1 if flooding frequency class is 2, and 0 otherwise), and ffreq3 (1 if flooding frequency class is 3) for the explanatory variables:

```
> x     <-data.frame(dist= meuse[,"dist"],
+                 ffreq2=ifelse(meuse$ffreq==2,1,0),
+                 ffreq3=ifelse(meuse$ffreq==3,1,0))
```

## 3.2. Model

The Moran eigenvectors, which are the basis functions used for spatial process modeling, are constructed as below:

```
> meig   <-meigen(coords)
 25/155 eigen-pairs are extracted
```

We first estimate the classical Gaussian regression model using the resf function. The error statistics including the restricted log-likelihood (rlogLik), Akaike Information Criterion (AIC), and Bayesian Information Criterion (BIC) are as follows:

```
> mod0   <-resf(y=y, x=x,meig=meig)
> mod0$e
                     stat
resid_SE      172.9239783
adjR2(cond)     0.7720376
rlogLik     -1032.7022760
AIC          2079.4045520
BIC          2100.7085278
```

Unfortunately, this model is not appropriate because of the non-Gaussianity of y. For non-negative explained variables like zinc concentration, user can specify y_nonneg = TRUE in the nongauss_y function. If it is specified, the explanatory variable y is assumed to be non-negative, and the Box-Cox transformation is applied:

```
> ng1    <-nongauss_y(y_nonneg=TRUE)
Box-cox transformation f() is applied to y to estimate
y ~ P( xb, par )    (or f(y,par)~N(xb, sig) )

 - P(): Distribution estimated through the transformation
 - xb : Regression term with fixed and random coefficients in b
        which is specified by resf or resf_vc function
 - par: Parameter estimating data distribution
```

The output ng1 is used as an output of the resf function to estimate a regression model with residual spatial dependence and Box-Cox transformation for y:

```
> mod1   <-resf(y=y,x=x, meig=meig, nongauss=ng1)
> mod1
Call:
resf(y = y, x = x, meig = meig, nongauss = ng1)

----Coefficients-----------------------------
              Estimate         SE    t_value        p_value
(Intercept)  3.1550749 0.01777841 177.466681 0.000000e+00
dist        -0.5160247 0.07024097  -7.346492 1.956835e-11
ffreq2      -0.1248181 0.01390843  -8.974277 2.664535e-15
ffreq3      -0.1318089 0.02119947  -6.217554 6.298492e-09

----Variance parameter-----------------------

Spatial effects (residuals):
                   (Intercept)
random_SE           0.09615471
Moran.I/max(Moran.I) 0.41327562

----Estimated probability distribution of y--------------
              Estimates
skewness        2.488325
excess kurtosis 7.972227
(Box-Cox parameter: -0.263962)

----Error statistics------------------------
                stat
resid_SE      0.0581104
adjR2(cond)   0.8453350
rloglik      -971.3835963
AIC          1958.7671926
BIC          1983.1145935

NULL model: lm( y ~ x )
   (r)loglik: -1083.605 ( AIC: 2177.211,  BIC: 2192.428 )
```

The resf_vc function is available when assuming SVCs. The estimated skewness, excess kurtosis, and the Box-Cox parameter confirm non-Gaussianity of the data. BIC of the model (1983.114), which considers residual spatial dependence, is considerably better than the ordinary linear regression model (2192.428). Accuracy of the model is confirmed.

In addition to the Box-Cox transformation, the SAL transformation can be iterated to estimate the most likely probability density function (PDF) behind y. The number of iterations is specified by an argument tr_num. We compare models with tr_num=1 (ng2) and tr_num=2 (ng3):

```
> ng2    <-nongauss_y(y_nonneg=TRUE, tr_num=1)
Box-Cox and 1 SAL transformations f() are applied to y to estimate
y ~ P( xb, par )   (or f(y,par)~N(xb, sig) )

 - P(): Distribution estimated through the transformation(s)
 - xb : Regression term with fixed and random coefficients in b
        which is specified by resf or resf_vc function
 - par: Parameters estimating data distribution
> ng3    <-nongauss_y(y_nonneg=TRUE, tr_num=2)
Box-Cox and 2 SAL transformations f() are applied to y to estimate
y ~ P( xb, par )   (or f(y,par)~N(xb, sig) )

 - P(): Distribution estimated through the transformation(s)
 - xb : Regression term with fixed and random coefficients in b
        which is specified by resf or resf_vc function
 - par: Parameters estimating data distribution
```

The following non-Gaussian models considering residual spatial dependence are estimated:

```
> mod2   <-resf(y=y, x=x,meig=meig, nongauss=ng2)
> mod3   <-resf(y=y, x=x,meig=meig, nongauss=ng3)
```

Model accuracy can be compared using the BIC (or AIC) value. Based on the BIC, mod2, which applies the Box-Cox transformation first and a SAL transformation after that, is the best model.

```
> mod2$e
                    stat
resid_SE      0.3976609
adjR2(cond)   0.8341787
rlogLik     -958.5890848
AIC         1937.1781696
BIC         1967.6124208
> mod3$e
                    stat
resid_SE      0.3996559
adjR2(cond)   0.8277254
rlogLik     -958.8305130
AIC         1945.6610260
BIC         1988.2689776
```

The estimated parameters are as follows:

```
> mod2  <-resf(y=y, x=x,meig=meig, nongauss=ng2)
> mod2
Call:
resf(y = y, x = x, meig = meig, nongauss = ng2)

----Coefficients------------------------------
             Estimate         SE    t_value       p_value
(Intercept)  1.2100004 0.11458914  10.559468  0.000000e+00
dist        -3.5209129 0.45132654  -7.801254  1.716405e-12
ffreq2      -0.7826159 0.09477395  -8.257712  1.421085e-13
ffreq3      -0.8259699 0.14323514  -5.766531  5.544422e-08

----Variance parameter------------------------

Spatial effects (residuals):
                    (Intercept)
random_SE             0.6035734
Moran.I/max(Moran.I)  0.3693597

----Estimated probability distribution of y--------------
               Estimates
skewness        1.717799
excess kurtosis 3.327901
(Box-Cox parameter: -0.2819055)

----Error statistics--------------------------
                    stat
resid_SE      0.3976609
adjR2(cond)   0.8341787
rlogLik     -958.5890848
AIC         1937.1781696
BIC         1967.6124208

NULL model: lm( y ~ x )
   (r)loglik: -1083.605 ( AIC: 2177.211,  BIC: 2192.428 )
```

The estimated PDF for y can be plotted as follows:

```
> plot(mod2$pdf,type="l")
```

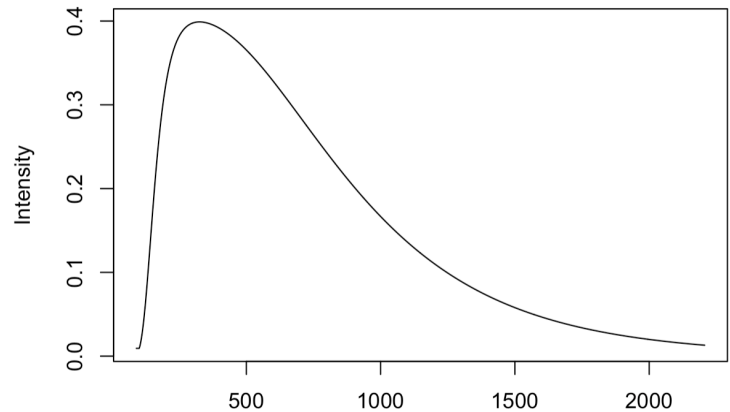

The estimated PDF is reasonably similar to the histogram of y.

While regression coefficients for transformed y is often difficult to interpret, the marginal effect of each explanatory variable $(dy_i/dx_{i,k})$ which quantifies the magnitude of change in $i$-th explained variable $(y_i)$ for one unit change in the $k$-th explanatory variable $(x_{i,k})$, is evaluated by using the coef_marginal function:

```
> coef_marginal(mod2)
Call:
coef_marginal(mod = mod2)

----Marginal effects from x (dy_i/dx_i) (summary)-------
 (Intercept)         dist              ffreq2             ffreq3
 Mode:logical   Min.   :-3832.79   Min.   :-851.94    Min.   :-899.13
 NA's:155       1st Qu.:-1858.62   1st Qu.:-413.13    1st Qu.:-436.01
                Median :-1173.63   Median :-260.87    Median :-275.32
                Mean   :-1195.47   Mean   :-265.73    Mean   :-280.45
                3rd Qu.: -368.10   3rd Qu.: -81.82    3rd Qu.: -86.35
                Max.   :  -98.77   Max.   : -21.95    Max.   : -23.17

Note: Medians are recommended summary statistics
```

For example, the median for ffreq2 suggests that areas with flooding frequency class 2 have 260.87 ppm smaller zinc concentration on median than other areas.

### 3.3. Spatial prediction and uncertainty analysis

The estimated model (mod2) is applied to spatially predict the zinc concentration on 3,103 grid points with 40 m × 40 m spacing (meuse.grid). Spatial coordinates (coords0) and the explanatory variables in the grids are used for the prediction:

```
> data(meuse.grid)
> coords0<-meuse.grid[,c("x","y")]
> x0      <-data.frame(dist= meuse.grid$dist,
+                     ffreq2=ifelse(meuse.grid$ffreq==2,1,0),
+                     ffreq3=ifelse(meuse.grid$ffreq==3,1,0))
```

The Moran eigenvectors at the prediction sites are generated using the meigen0 function:

```
> meig0   <-meigen0(meig=meig, coords0=coords0)
> pres    <-predict0(mod=mod2,x0=x0,meig0=meig0, compute_quantile = TRUE)
```

The spatial prediction is performed using the predict0 function. If compute_quantile=TRUE, quantiles for the predicted values are evaluated based on the PDF estimated in Section 1.2:

```
> meig0   <-meigen0(meig=meig, coords0=coords0)
> pres    <-predict0(mod=mod2,x0=x0,meig0=meig0, compute_quantile = TRUE)
```

The outputs are as follows:

```
> pres$pred[1:2,]
      pred pred_transG pred_transG_se   xb  sf_residual
1 916.2723    1.191011      0.4128080 1.21 -0.018989791
2 923.0430    1.201812      0.4132363 1.21 -0.008188592
```

The output includes the predicted values on the original scale (pred), the predicted value on the transformed scale (pred_transG), and the standard error (pred_transG_se). The estimated quantiles for the predicted values are displayed as follows:

```
> pres$pred_quantile[1:2,]
     q0.01    q0.025     q0.05      q0.1      q0.2      q0.3      q0.4      q0.5
1 414.9931  482.7518  544.3806  618.7869  714.1259  786.9783  852.3321  916.2723
2 419.2256  487.3734  549.2768  624.0092  719.8005  793.0283  858.7385  923.0430
      q0.6      q0.7      q0.8      q0.9     q0.95    q0.975     q0.99
1  983.2187 1058.455  1151.664  1291.096  1416.133  1532.610  1678.346
2  990.3854 1066.082  1159.881  1300.232  1426.124  1543.421  1690.210
```

To map the outputs, pred, pred_transG, pred_transG_se, quantiles for the predicted values (pred_quantile) are summarized into a data.frame object. As a measure of uncertainty, the length of the 95 % confidence interval for the predicted value (len95) is added. Besides, predicted values of a regression kriging, which is widely used for spatial prediction, is also added (kpred). The data.frame object is then converted to a sf object for mapping:

```
> pred    <- data.frame(coords0,pred=pres$pred[,"pred"],
+                       len95=pres$pred_quantile$q0.975 - pres$pred_quantile$q0.025,
+                       pred_transG=pres$pred[,"pred_transG"],
+                       pred_transG_se=pres$pred[,"pred_transG_se"],
+                       pres$pred_quantile,
+                       kpred=exp(kres$krige_output$var1.pred))
> coordinates(pred)<-c("x","y")
> pred_sf <-st_as_sf(pred)
```

Our prediction result (pred) and the kriging-based prediction result (kpred) are quite similar:

```
> plot(pred_sf[,c("pred","kpred")], pch=20, axes=TRUE, key.pos = 1)
```

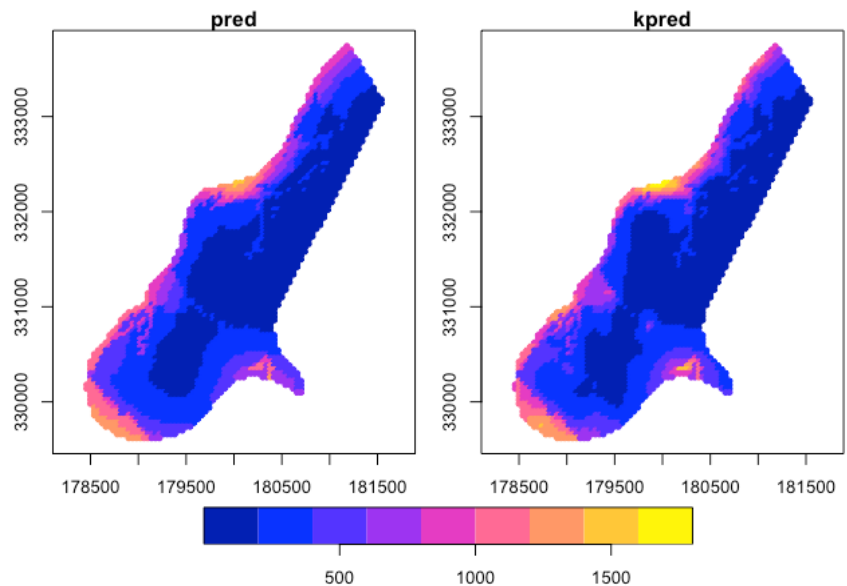

As shown in the maps below showing the 2.5%, 50%, and 97.5% quantiles, the predicted values have larger uncertainty in the north area that faces the river Meuse:

```
> plot(pred_sf[,c("q0.025","q0.5","q0.975")], pch=20, axes=TRUE, key.pos = 1)
```

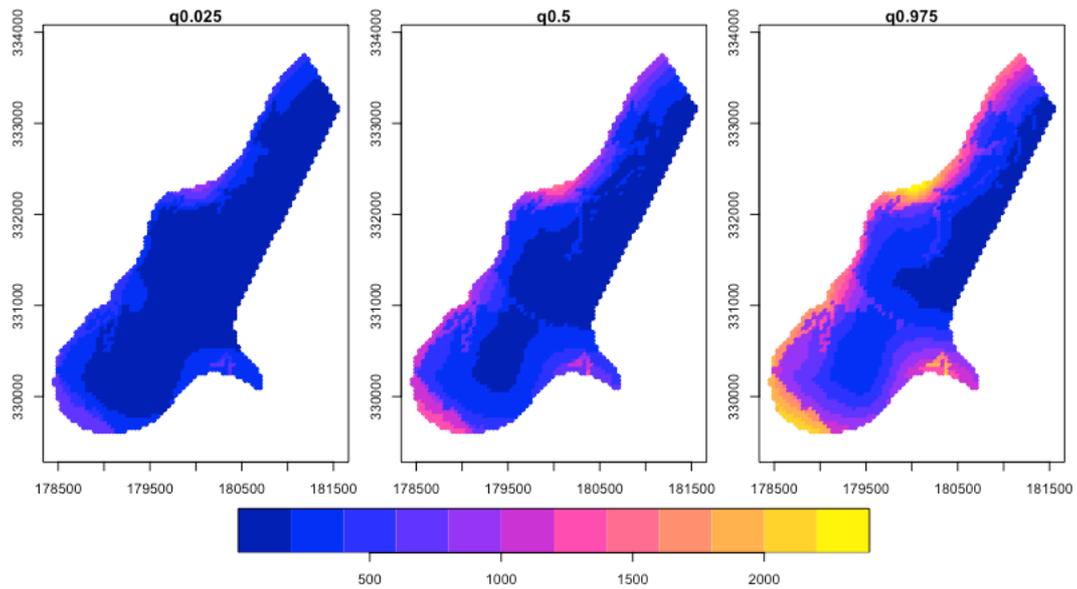

The map below is the length of the 95 % confidence interval (len95), which is another way to visualize the uncertainty in the original scale:

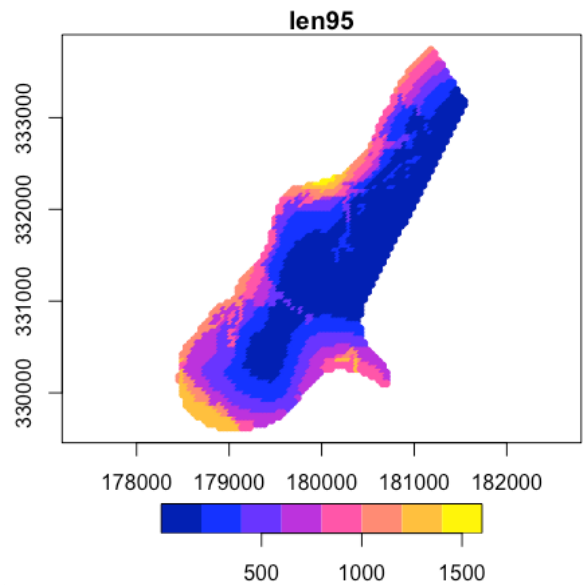

We can also visualize the predicted values in the transformed/normalized scale:

```
> plot(pred_sf[,"pred_transG"], pch=20, axes=TRUE, key.pos = 1)
```

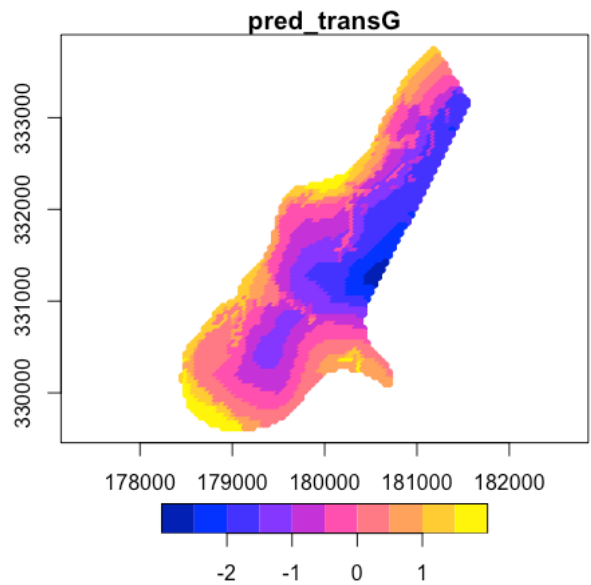

As shown below, in the transformed scale, the predictive errors are large in the eastern central area where the samples are relatively limited (but, as far as we see the maps for len95 or the quantiles, this error has little impact in the original scale as a result of the rescaling/transformation to the real scale).

```
> plot(pred_sf[,"pred_transG_se"], pch=20, axes=TRUE, key.pos = 1)
```

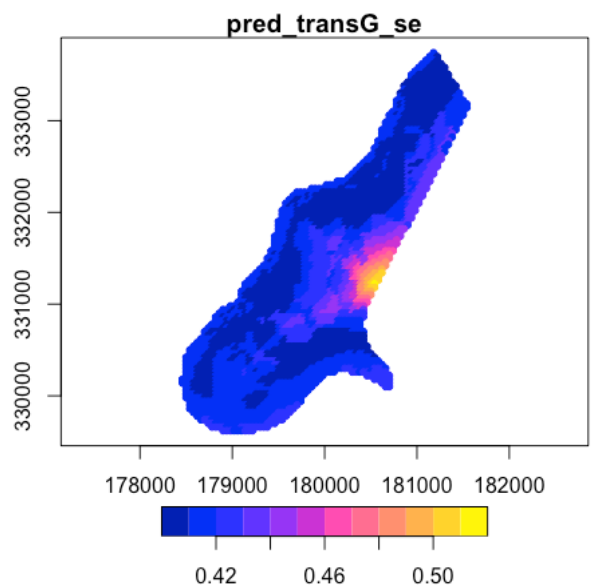

### 3.4. Limitation

The Moran eigenvector approach provides a kind of low rank approximation for spatial process modeling (just like fixed rank kriging and predictive process modeling; see Sun et al., 2012). While the modeling accuracy is good enough in many cases, it can provide overly smoothed spatial prediction result for very large samples (e.g., $N > 10,000$; see, Stein, 2014). For spatial prediction using large samples, it should be used with caution (at least, this approach is still useful even in such a case to understand underlying map patterns computationally efficiently).

## 4. Example 3: Non-Gaussian spatial hedonic analysis

This section demonstrates the importance of considering non-Gaussianity in hedonic housing price analysis. Gaussian and non-Gaussian spatial varying coefficient (SVC) models are use in this section.

### 4.1. Data

This section uses the spdep, sf, spmoran packages:

```
> library(spdep);library(sf);library(spmoran)
```

This section analyzes the housing data for 506 census tracts in Boston in 1970. Explained variable (y) is the median housing value in USD 1000's (CMEDV). The explained variables whose coefficients are allowed to vary over space (x), those whose coefficients are assumed constant (xconst), and spatial coordinates (coords) are used in this analysis:

```
> data(boston)
> y       <- boston.c[, "CMEDV"]
> x       <- boston.c[,c("CRIM", "AGE")]
> xconst  <- boston.c[,c("ZN","DIS","RAD","NOX",  "TAX","RM", "PTRATIO", "B")]
> coords  <- boston.c[,c("LON","LAT")]
```

Moran eigenvectors are extracted as follows:

```
> meig    <- meigen(coords=coords)
 55/506 eigen-pairs are extracted
```

### 4.2. Model

This section considers three transformations functions:

```
> ng1      <- nongauss_y(y_nonneg=TRUE)
Box-cox transformation f() is applied to y to estimate
y ~ P( xb, par )    (or f(y,par)~N(xb, sig) )

 - P(): Distribution estimated through the transformation
 - xb : Regression term with fixed and random coefficients in b
        which is specified by resf or resf_vc function
 - par: Parameter estimating data distribution

> ng2      <- nongauss_y(y_nonneg=TRUE,tr_num=1)
Box-Cox and 1 SAL transformations f() are applied to y to estimate
y ~ P( xb, par )    (or f(y,par)~N(xb, sig) )

 - P(): Distribution estimated through the transformation(s)
 - xb : Regression term with fixed and random coefficients in b
        which is specified by resf or resf_vc function
 - par: Parameters estimating data distribution

> ng3      <- nongauss_y(y_nonneg=TRUE,tr_num=2)
Box-Cox and 2 SAL transformations f() are applied to y to estimate
y ~ P( xb, par )    (or f(y,par)~N(xb, sig) )

 - P(): Distribution estimated through the transformation(s)
 - xb : Regression term with fixed and random coefficients in b
        which is specified by resf or resf_vc function
 - par: Parameters estimating data distribution
```

Although ng3 is the most flexible, it can lead to overfitting. To identify the best model, the Gaussian SVC model (mod0) and non-Gaussian SVC models (mod1, mod2, mod3) are fitted, and their BIC values are compared:

```
> mod0     <- resf_vc(y=y,x=x, x_nvc=TRUE,xconst=xconst,meig=meig )
> mod1     <- resf_vc(y=y,x=x, x_nvc=TRUE,xconst=xconst,meig=meig, nongaus=ng1 )
> mod2     <- resf_vc(y=y,x=x, x_nvc=TRUE,xconst=xconst,meig=meig, nongaus=ng2 )
> mod3     <- resf_vc(y=y,x=x, x_nvc=TRUE,xconst=xconst,meig=meig, nongaus=ng3 )
```

The resulting BICs are 3110.5 (mod0), 2950.5 (mod1), 2901.6 (mod2), 2931.4 (mod3), and 3178.4 for the ordinary liner regression model. mod2, which applies the Box-Cox transformation and a SAL transformation is selected as the best model.

The parameters estimated from mod2 are as follows:

```
> mod2
Call:
resf_vc(y = y, x = x, xconst = xconst, x_nvc = TRUE, meig = meig,
    nongauss = ng2)

----Spatially and non-spatially varying coefficients on x (summary)----

Coefficient estimates:
  (Intercept)            CRIM                  AGE
 Min.   :-0.02244   Min.   :-0.2740242   Min.   :-0.018914
 1st Qu.:-0.02244   1st Qu.:-0.0599508   1st Qu.:-0.010591
 Median :-0.02244   Median :-0.0322745   Median :-0.007599
 Mean   :-0.02244   Mean   :-0.0329763   Mean   :-0.007425
 3rd Qu.:-0.02244   3rd Qu.: 0.0004135   3rd Qu.:-0.004354
 Max.   :-0.02244   Max.   : 0.1070968   Max.   : 0.005453

Statistical significance:
                       Intercept CRIM AGE
Not significant              506  410 117
Significant (10% level)        0   18  24
Significant ( 5% level)        0   19  52
Significant ( 1% level)        0   59 313
----Constant coefficients on xconst---------------------------
           Estimate           SE    t_value       p_value
ZN       0.002027180 0.0011645284   1.740773  8.243151e-02
DIS     -0.131266652 0.0237841152  -5.519089  5.869668e-08
RAD      0.052234354 0.0085592354   6.102689  2.312320e-09
NOX     -3.124557004 0.4565365150  -6.844046  2.632916e-11
TAX     -0.001635874 0.0003135737  -5.216872  2.823456e-07
RM       0.506995252 0.0296312602  17.110148  0.000000e+00
PTRATIO -0.056300954 0.0135694652  -4.149092  4.017337e-05
B        0.002452484 0.0002849729   8.606026  0.000000e+00

----Variance parameters----------------------------------

Spatial effects (coefficients on x):
                     (Intercept)        CRIM         AGE
random_SE           3.398454e-06 0.12892890 0.007028641
Moran.I/max(Moran.I) 4.631293e-01 0.05171784 0.273869153

Non-spatial effects (coefficients on x):
                CRIM AGE
random_SE 0.003807227   0

----Estimated probability distribution of y--------------
                 Estimates
skewness          1.200526
excess kurtosis   1.765607
(Box-Cox parameter: 1.691544)

----Error statistics-----------------------------------
                stat
resid_SE       0.3303358
adjR2(cond)    0.8881671
rlogLik       -1382.3284183
AIC            2808.6568365
BIC            2901.6406432

NULL model: lm( y ~ x + xconst )
   (r)loglik: -1551.857 ( AIC: 3127.715,  BIC: 3178.433 )
```

The "Estimated probability distribution of y" section suggests that the data is positively skewed (skewness > 0) and fat tail (excess kurtosis > 0). The estimated probability density distribution can be visualized as follows:

```
> plot(mod2$pdf,type="l")
```

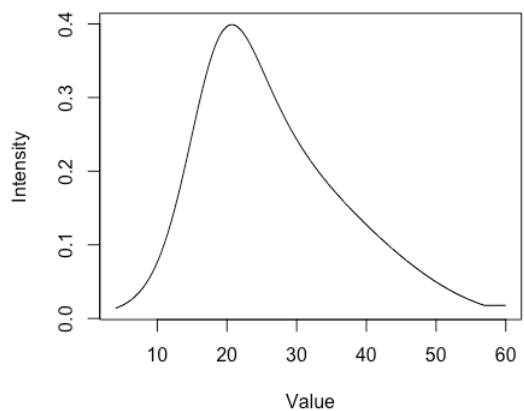

Marginal effect of each explanatory variable ($dy_i/dx_{i,k}$) which quantifies the magnitude of change in $i$-th explained variable ($y_i$) for one unit change in the $k$-th explanatory variable ($x_{i,k}$), is evaluated using the coef_marginal function if the resf function is used while the coef_marginal_vc function if the resf_vc function is used like our case:

```
> coef_marginal_vc(mod2)
Call:
coef_marginal_vc(mod = mod2)

----Marginal effects from x (dy_i/dx_i) (summary)----
 (Intercept)          CRIM                 AGE
 Mode:logical   Min.   :-3.186702   Min.   :-0.32519
 NA's:506       1st Qu.:-0.485693   1st Qu.:-0.08374
                Median :-0.240681   Median :-0.05859
                Mean   :-0.285872   Mean   :-0.05956
                3rd Qu.: 0.003341   3rd Qu.:-0.03813
                Max.   : 1.443992   Max.   : 0.09132
----Marginal effects from xconst (dy_i/dx_i)(summary)----
       ZN                DIS               RAD              NOX
 Min.   :0.01145   Min.   :-2.7675   Min.   :0.2950   Min.   :-65.88
 1st Qu.:0.01209   1st Qu.:-1.2022   1st Qu.:0.3114   1st Qu.:-28.62
 Median :0.01402   Median :-0.9079   Median :0.3613   Median :-21.61
 Mean   :0.01841   Mean   :-1.1922   Mean   :0.4744   Mean   :-28.38
 3rd Qu.:0.01857   3rd Qu.:-0.7826   3rd Qu.:0.4784   3rd Qu.:-18.63
 Max.   :0.04274   Max.   :-0.7412   Max.   :1.1013   Max.   :-17.64
       TAX               RM              PTRATIO             B
 Min.   :-0.034490   Min.   : 2.863   Min.   :-1.1870   Min.   :0.01385
 1st Qu.:-0.014982   1st Qu.: 3.023   1st Qu.:-0.5156   1st Qu.:0.01462
 Median :-0.011315   Median : 3.507   Median :-0.3894   Median :0.01696
 Mean   :-0.014857   Mean   : 4.605   Mean   :-0.5113   Mean   :0.02227
 3rd Qu.:-0.009753   3rd Qu.: 4.643   3rd Qu.:-0.3357   3rd Qu.:0.02246
 Max.   :-0.009238   Max.   :10.689   Max.   :-0.3179   Max.   :0.05171

Note: Medians are recommended summary statistics
```

For example, the median of CRIM suggests that, on median, housing price decreases 0.24 (1,000 USD) for every 1.0 increase of CRIM (per capita crime rate).

The estimated SVCs on x (CRIM, AGE, and Intercept) can be plotted using the plot_s function. For example, the SVC on CRIM, which is the first column of x, is mapped as follows:

```
> plot_s(mod2,1)
```

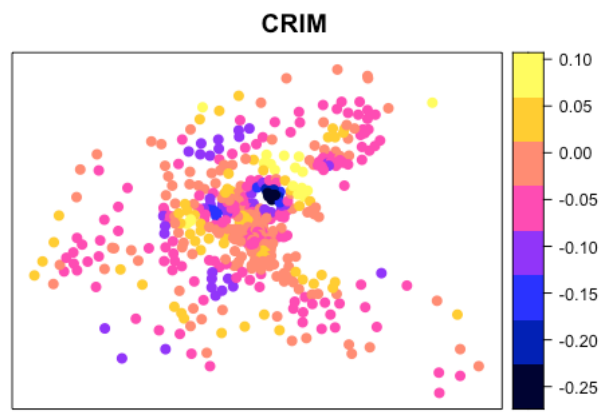

The output suggests the strong negative impact of CRIM in the central area. An argument pmax is useful to display statistically significant coefficients only. For example, here is the code to display the coefficients that are statistically significant at the 5 % level:

```
> plot_s(mod2,1,pmax=0.05)
```

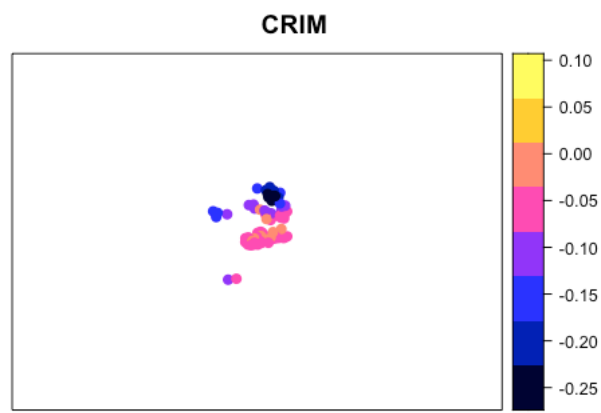

This map demonstrates that the crime rate has statistically significant negative impact on housing price only in the central area. Alternatively, the SVCs can be plotted using the sf package as follows:

```
> boston.tr <- boston.tr0[order(boston.tr0$TOWNNO),1:8]
> b_est    <- mod2$b_vc
> boston.tr <- cbind(boston.tr, b_est)
> names(boston.tr)
 [1] "poltract"   "TOWN"       "TOWNNO"
 [4] "TRACT"      "LON"        "LAT"
 [7] "MEDV"       "CMEDV"      "X.Intercept."
[10] "CRIM"       "AGE"        "geometry"
> plot(boston.tr[,"CRIM"],axes=TRUE,lwd=0.1, key.pos = 1)
```

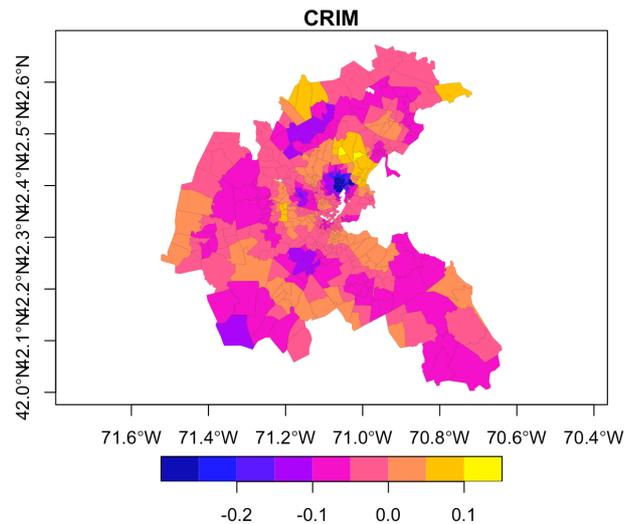